\documentclass{article}
\usepackage{graphicx}

\begin{document}
\date{}
\def\be{\begin{equation}}
\def\ee#1{\label{#1}\end{equation}}

\title{Acceleration field of a Universe modeled
as a mixture of scalar and matter fields}
\author{Gilberto  M. Kremer\thanks{E-mail: 
kremer@fisica.ufpr.br}$\,$ and Daniele S. M. Alves\thanks{E-mail: 
dsma01@fisica.ufpr.br}
\\Departamento de F\'\i sica, 
Universidade Federal do Paran\'a,\\
Caixa Postal 19044, 81531-990 Curitiba, Brazil}
\maketitle

\begin{abstract}
A model of the Universe  
as a mixture of a scalar (inflaton or rolling 
tachyon from the string theory) and a matter field (classical particles)
is analyzed. 
The particles are created at the expense 
of the gravitational energy through an irreversible process whereas
the scalar field is supposed to
interact only with itself and to be minimally coupled with the gravitational
field. The irreversible processes of particle creation are related to 
the non-equilibrium pressure within the framework of the
extended (causal or second-order) thermodynamic theory. 
The scalar field (inflaton or tachyon) is described by an 
exponential potential density added by a parameter which represents its 
asymptotic value 
and can be interpreted as  the vacuum energy.
This model can simulate  three phases of the acceleration field
of the Universe, namely,
(a) an inflationary epoch with a positive acceleration 
followed by a decrease  of the acceleration field towards zero, (b) 
a past decelerated period where the acceleration field decreases to 
a maximum negative value followed by an increase towards zero, and 
(c) a present accelerated epoch. For the energy densities there exist
also three distinct epochs which begin with a scalar field dominated period 
followed by a matter field dominated epoch and coming back to a scalar field 
dominated phase.
\end{abstract}

\section{Introduction}

The solution of the problems   of flatness, 
horizon and unwanted relics  by the inflationary theory on the one hand
and the  recent measurements of the anisotropy of the 
cosmic microwave background 
and of the  type 
Ia supernova SN 1997ff red-shift indicating that the Universe is flat with a 
present positive acceleration and a past 
decelerating period on the other hand, allows us to classify  the
evolution of the acceleration field of the Universe  according to three 
major epochs, beginning with an accelerated inflationary period 
following a past decelerated epoch  and leading back to a present 
accelerated phase.

The early and present accelerated periods are dominated
by a scalar field 
whereas the past decelerated phase is dominated by a matter
field. In inflationary cosmology the scalar field is normally 
represented by the inflaton (see, e.g., the works~\cite{Guth,Linde,AS}) 
but recently a rolling tachyon -- which  has arisen from string 
theory~\cite{Sen1,Sen2} --
was also considered as another candidate for the description of 
the accelerated phases of the Universe (see, e.g., the 
works~\cite{tach1,tach3,tach4,tach6}).\footnote{Recently a model 
for the accelerated phase 
of the Universe was analyzed in the works~\cite{Cap,K2} where 
the  van der Waals equation of state plays the role of the scalar field.}
The inflationary period comes to an end when the scalar field has rolled to its
potential minimum and begins to oscillate about it. By this time, the Universe 
is very cold and void. Some mechanism is then required to account for particle
creation, that will fill and heat the Universe to allow for the standard hot
Big-Bang evolution to take place. The most common mechanism adopted is the 
decay of the inflaton field in other particles, and it is known as reheating 
process. It goes without saying that this requires the coupling of the inflaton
to other fields. However, the inflaton needs to be weakly coupled in order to 
inflation to work well. In view of that, in the present work we will make 
another approach and ignore the coupling of the scalar field to other fields. 
It is known from quantum field theory in curved space-time (see 
e.g.~\cite{QFT}), that
matter constituents may be produced quantum-mechanically in the framework of
Einstein's equations. The energy of the produced particles is extracted from 
the gravitational field. Our procedure will then be to take this fact into 
account by a phenomenological description
-- first proposed, to the best of our knowledge, by Prigogine et 
al~\cite{PG} --  in which the particles are created during the
evolution of the Universe at the expense 
of the gravitational energy through an irreversible process (see also the 
works~\cite{Z,KD0,KD1,K1}).

In the present work we have considered 
that the irreversible processes of particle production are related to 
a non-equilibrium pressure with an evolution equation coming from the extended 
(causal or second-order) thermodynamic 
theory (see e.g.~\cite{KD0,KD1,K1,Be,RP,CJ,CH,Zim}). 
 Moreover, since the scalar field potential does not require a minimum about
which the inflaton will oscillate and decay in other particles, we shall adopt 
an exponential potential density added by a parameter which represents its 
asymptotic value and can be interpreted as  the vacuum 
energy~\cite{SS,Car,PR,Pad}.  

We have shown, among other results, that a mixture of a scalar field (inflaton
or tachyon) and a matter field can simulate the three phases of the Universe 
related to its acceleration field, namely,
(a) an inflationary epoch with a positive acceleration 
followed by a decrease  of the acceleration field towards zero, (b) 
a past decelerated period where the acceleration field decreases to 
a maximum negative value followed by an increase towards zero, and 
(c) a present accelerated epoch. For the energy densities there exist
also three distinct epochs which begin with a scalar field dominated period 
followed by a matter field dominated epoch and coming back to a scalar field 
dominated phase.  These results can  also be obtained by choosing  other 
types of potential densities for the scalar field (inflaton or tachyon). 

The work is organized as follows.   
In section 2 a  system of three coupled 
differential equations for the scalar (inflaton or tachyon), 
acceleration and  non-equilibrium pressure fields
is determined. 
The solutions of the systems of coupled differential equations obtained
in section 2 are  found 
in section 3 for given initial conditions and for a given potential 
energy density of the scalar field. We close the work with a discussion 
of the  results obtained 
in section 3.  Units have been chosen so that $c=\hbar=k=1$.

\section{Field Equations}

Let us consider a homogeneous, isotropic and spatially flat Universe 
modeled as a mixture
of a scalar field and a matter field. The scalar field (inflaton or tachyon) 
represents a hypothetical
particle  while the matter field refers to the  
classical particles which are created at the expense of the gravitational 
energy.

For this model of the Universe the energy-momentum tensor of the mixture 
is written as
\be
T^{\mu\nu}=(\rho+p+\varpi)U^\mu U^\nu-
(p+\varpi)g^{\mu\nu},
\ee{1}
where the pressure and the energy density of the mixture are given in terms 
of the corresponding quantities for its constituents by  
$p=p_s+p_m$ and  $\rho=\rho_s+\rho_m$, with the indexes $s$ and 
$m$ denoting the scalar and the matter fields, respectively. We shall 
adopt the following convention:
(a) $s=\phi$ for the inflaton field and (b) $s=\varphi$
for the tachyon field.  Furthermore,
$U^\mu$ (such that $U^\mu U_\mu=1$) is the four-velocity, $g_{\mu\nu}$ denotes 
the metric tensor with signature $(+---)$ whereas
 $\varpi$ refers to the non-equilibrium pressure. The non-equilibrium
pressure  is responsible  for the irreversible processes of
particle production~\cite{Z,KD0} during the 
evolution of the Universe.

The conservation law of the 
energy-momentum tensor ${T^{\mu\nu}}_{;\nu}=0$ follows from Einstein's
field equations and the
Bianchi identities. In a comoving frame described
by the Robertson-Walker metric it 
leads to the balance equation for the energy 
density of the mixture, i.e.,
\be
\dot\rho+3H(\rho+p+\varpi)=0,
\ee{2}
where the quantity $H=\dot a(t)/a(t)$ denotes the Hubble parameter,
$a(t)$ is the cosmic scale factor and the over-dot  
refers to  differentiation with respect to time $t$.

The equation which
connects the Hubble parameter  
with the energy density of the mixture is the Friedmann equation 
which -- in a spatially
flat Universe described by the Robertson-Walker metric -- is written as
\be
H^2={8\pi G\over3}\rho,
\ee{9}
where $G$ is the gravitational constant. 

\subsection{Inflaton field}

First we shall analyze the case where the scalar field is represented by the
inflaton field $\phi(x^\mu)$, which is  described by the Lagrangian density 
\be 
{\cal L}_\phi={1\over 2}\partial_\mu\phi\partial^\mu\phi-V(\phi),
\ee{3}
where $V(\phi)$ denotes the potential density of the inflaton field. The 
identification of the inflaton with a perfect fluid -- i.e., by requiring 
that the inflaton interacts only with itself and it is minimally 
coupled with the gravitational field --
allows us to write its energy-momentum tensor as
\be
T^{\mu\nu}_\phi=(\rho_\phi+p_\phi)U^\mu U^\nu-p_\phi g^{\mu\nu}=
\partial^\mu\phi\partial^\nu\phi-{\cal L}_\phi g^{\mu\nu}.
\ee{4}
The last equality above  is a consequence of 
Noether's theorem.

By considering a homogeneous inflaton field and a comoving frame 
one  can obtain from equations 
(\ref{3}) and (\ref{4}) the relationships 
\be
\cases{\rho_\phi={1\over 2}\dot\phi^2+V(\phi),\cr 
p_\phi={1\over 2}\dot\phi^2-V(\phi),}
\ee{5}
which connect the energy density $\rho_\phi$ and the
pressure $p_\phi$ of the inflaton to its kinetic and potential energies.
  
The time evolution equation of the inflaton field 
follows from the Euler-Lagrange
equation which, in the homogeneous case, reads
\be
\ddot\phi+3H\dot \phi+V'(\phi)=0.
\ee{6}
Above, the prime denotes differentiation with respect to $\phi$.

The evolution equation of the energy density of the inflaton field 
decouples from that of the matter field, since the differentiation 
of (\ref{5})$_1$ with respect to time by taking into account the 
equation (\ref{6}) leads to
\be
\dot\rho_\phi+3H(\rho_\phi+p_\phi)=0.
\ee{7}
Hence the evolution equation 
of the energy density of the matter field can be written as
\be
\dot\rho_m+3H(\rho_m+p_m)=-3H\varpi.
\ee{8}
thanks to (\ref{2}) and (\ref{7}). We can interpret 
the term $-3H\varpi$ in the above equation
as the energy  production rate of the matter field 
(see e.g.,~\cite{KD0,KD1,K1}). In a previous work~\cite{KD0} one of the 
authors have calculated the energy-momentum pseudo-tensor of the gravitational
field and found $T^{00}_G=-3H^2/(8\pi G)$. If we identify $T^{00}_G$ with 
the energy density of the gravitational field $\rho_G$, we can regard the term 
$3H\varpi$ as the 
energy production rate of the gravitational field:
\be
\dot\rho_G+3H(\rho_G-p_\phi-p_m)=3H\varpi,
\ee{8a}
that is, there is an irreversible energy flow from the gravitational field to
matter creation.

The matter field is supposed to obey a barotropic equation of state 
$p_m=w_m\rho_m$ where the 
coefficient $w_m$ may assume values in the range between 
$0\leq w_m\leq1$. Some values for
this coefficient are: (a) $w_m=0$ for dust or pressure-less fluid; 
(b) $w_m=1/3$
 for radiation; (c) $w_m=2/3$ for non-relativistic matter and 
(d) $w_m=1$ for 
stiff matter or Zel'dovich fluid.

The equation which gives the time evolution of the cosmic scale factor 
can be obtained from  differentiation of
the Friedmann equation  (\ref{9})  with respect to time and 
elimination of $\dot\rho$ from the resulting equation,
by using the balance equation for the energy density of the mixture (\ref{2}). 
Hence, it follows
\be
\dot H+{3\over 2}(w_m+1)H^2=4\pi G\left[(w_m-1){\dot\phi^2\over2}+
(w_m+1)V(\phi)-\varpi\right],
\ee{10}
thanks to (\ref{5}), (\ref{9}) and to the barotropic equation of state 
of the matter field.

Equation (\ref{10}) refers to a differential equation for the
cosmic scale factor $a(t)$ that depends on the non-equilibrium pressure 
$\varpi(t)$
and on the potential density of the inflaton field $V(\phi)$.
If we know a relationship between $\varpi(t)$ and $a(t)$ for a given
$V(\phi)$  it would be possible to find
a solution of the system of differential equations (\ref{6}) and 
(\ref{10}) for the time evolution of
the cosmic scale factor and for the inflaton field. 
Here we shall consider that the non-equilibrium pressure
obeys -- within the framework of extended (causal or second-order) 
thermodynamic theory -- the linearized evolution equation
\footnote{One is referred 
to~\cite{CK} for a derivation of the  evolution equation (\ref{11})
within the framework of the Boltzmann equation.} 
 \be
\varpi+\tau\dot\varpi=-3\eta H.
\ee{11}
Above,  the coefficient of bulk viscosity  $\eta$ and the characteristic time 
$\tau$ are considered  as functions of the energy 
density of the mixture $\rho$, i.e., $\eta=\alpha\rho$,
and $\tau={\eta/ \rho}$ where $\alpha$ is a constant 
(see e.g.~\cite{KD0,KD1,K1,Be}).

Now we have a system of three differential equations (\ref{6}), (\ref{10})
and (\ref{11}), and in order to solve it we introduce the 
dimensionless quantities
\be\cases{
H\equiv {H/H_0},\quad t\equiv tH_0,\quad
\varpi\equiv \varpi[ {8\pi G/(3H_0^2)}],\cr\alpha\equiv\alpha H_0,\quad
V\equiv V[{8\pi G/(3H_0^2)}],\quad \phi\equiv \phi\sqrt{8\pi G/3}.}
\ee{12}
Above, the Hubble parameter $H_0$ at $t=0$ 
(by adjusting clocks) is related to the energy density $\rho_\phi^0$ 
 of the inflaton field at $t=0$
by $H_0=\sqrt{8\pi G \rho_\phi^0/3}$, since we have assumed that at $t=0$
the energy density of the matter field vanishes ($\rho_m^0=0$)
and $\dot \phi(0)=0$.

With respect to the dimensionless quantities (\ref{12}) 
the system of differential equations (\ref{6}), (\ref{10}) and (\ref{11})
reads
\be
\ddot\phi+3H\dot \phi+V'(\phi)=0,
\ee{13}
\be
\dot H+{3\over 2}(w_m+1)H^2={3\over2}\left[(w_m-1){\dot\phi^2\over2}+
(w_m+1)V(\phi)-\varpi\right],
\ee{14}
\be
\varpi+\alpha\dot\varpi=-3\alpha H^3.
\ee{15}

The time evolution of the inflaton field $\phi(t)$, of the cosmic scale factor
$a(t)$ and of the non-equilibrium pressure $\varpi(t)$ 
can be determined from  the system of differential equations (\ref{13}) 
through (\ref{15}) 
once: (a) a potential density of the inflaton field $V(\phi)$ is chosen; (b) 
initial conditions at $t=0$ (by adjusting clocks) are given for the 
cosmic scale
factor $a(0)$ and its derivative $\dot a(0)$, for the inflaton field 
$\phi(0)$ and 
its derivative $\dot\phi(0)$ and for
the non-equilibrium pressure $\varpi(0)$; (c) values for coefficient $w_m$ --
related to the barotropic equation of state of the matter field -- 
and for the coefficient
$\alpha$ -- related to the irreversible processes of particle production
-- are selected. 

From the knowledge of the  fields $\phi(t)$, $a(t)$ and 
$\varpi(t)$ it is possible to determine the time evolution of the energy 
densities of the inflaton and matter fields from 
\be
{\rho_\phi\over\rho_\phi^0}={1\over 2}\dot\phi^2+V(\phi),\qquad 
{\rho_m\over\rho_\phi^0}=H^2-\left[{1\over 2}\dot\phi^2+V(\phi)\right],
\ee{16}
thanks to (\ref{5})$_1$ and (\ref{9}).

\subsection{Tachyon field}

Recently, Sen~\cite{Sen1,Sen2} has shown from a string theory 
that a rolling tachyon can be described by the Lagrangian density 
\be
{\cal L}_\varphi=-V(\varphi)\sqrt{1-\partial_\mu\varphi\partial^\mu\varphi},
\ee{t1}
where $V(\varphi)$ denotes the potential density of the tachyon field 
$\varphi(x^\mu)$.
The corresponding Born-Infeld action for the tachyon field reads
\be
S=\int d^4x\sqrt{-g}\left[{R\over 16\pi G}-V(\varphi)\sqrt{1-
\partial_\mu\varphi
\partial^\mu\varphi}\right].
\ee{t2}
In the above equation $R$ represents the curvature scalar.

From the action (\ref{t2}) it follows the energy-momentum tensor of the tachyon
field
\be
T^{\mu\nu}_\varphi=(\rho_\varphi+p_\varphi)U^\mu U^\nu-p_\varphi g^{\mu\nu},
\ee{t3}
where the energy density $\rho_\varphi$, the pressure $p_\varphi$  
and the four-velocity $U^\mu$ are identified as
\be
\cases{\rho_\varphi={V(\varphi)\over\sqrt{1-\partial_\mu\varphi\partial^\mu
\varphi}},\cr
p_\varphi=-V(\varphi)\sqrt{1-\partial_\mu\varphi\partial^\mu\varphi},\cr
U^\mu={\partial^\mu\varphi\over\sqrt{\partial_\nu
\varphi\partial^\nu\varphi}}.}
\ee{t4}

The field equation for the tachyon is obtained from the Euler-Lagrange 
equation and reads
\be
{(\partial_\mu\varphi)}_{;\nu}\left[g^{\mu\nu}+{\partial^\mu
\varphi\partial^\nu\varphi\over
1-\partial_\sigma\varphi\partial^\sigma\varphi}\right]
+{V'(\varphi)\over V(\varphi)}=0.
\ee{t5}

For a homogeneous tachyon field in a spatially flat Robertson-Walker metric, 
equations (\ref{t4})$_{1,2}$ and (\ref{t5}) reduce to
\be
\rho_\varphi={V(\varphi)\over\sqrt{1-\dot\varphi^2}},
\qquad p_\varphi=-V(\varphi)\sqrt{1-\dot\varphi^2},
\ee{t6}
\be
{\ddot\varphi\over 1-\dot\varphi^2}+3H\dot \varphi+{V'(\varphi)\over 
V(\varphi)}=0,
\ee{t7}
respectively, whereas the four-velocity (\ref{t4})$_{3}$ becomes that of 
a comoving frame.

The differentiation of equation (\ref{t6})$_1$ with respect to time leads to 
the evolution equation of the energy density of the tachyon field, i.e.,
\be
\dot\rho_\varphi+3H(\rho_\varphi+p_\varphi)=0,
\ee{t8}
thanks to (\ref{t6}) and (\ref{t7}). Hence, the energy density of 
the matter field 
decouples from the corresponding equation for the tachyon field, has the 
same expression as that given by (\ref{8}) and can be interpreted in the 
same manner as in the previous section. 

Furthermore, from the differentiation of the Friedmann equation (\ref{9}) 
with respect to time and by taking into account the barotropic equation of 
state for the matter field, it follows the dimensionless equation
\be
\dot H+{3\over 2}(w_m+1)H^2={3\over2}\left[{V(\varphi)\over
\sqrt{1-\dot\varphi^2}}(w_m+1
-\dot\varphi^2)-\varpi\right],
\ee{t9}
which relates the time evolution of the cosmic scale factor with the 
non-equilibrium pressure and the potential density of the tachyon field.
The dimensionless quantities in this case are
\be\cases{
H\equiv {H/H_0},\quad t\equiv tH_0,\quad
\varpi\equiv \varpi[ {8\pi G/(3H_0^2)}],\cr\alpha\equiv\alpha H_0,\quad
V\equiv V[{8\pi G/(3H_0^2)}],\quad \varphi\equiv \varphi H_0,}
\ee{t10}
where the Hubble parameter $H_0$ at $t=0$ 
(by adjusting clocks) is connected with the energy density $\rho_\varphi^0$ 
 of the tachyon field at $t=0$
by $H_0=\sqrt{8\pi G \rho_\varphi^0/3}$. Here we have also assumed 
that at $t=0$
the energy density of the matter field vanishes ($\rho_m^0=0$) and 
$\dot \varphi(0)=0$.

The system of differential equations we have to solve now consists of
the evolution equations:
(a)  for the non-equilibrium pressure (\ref{15});
(b) for the tachyon field (\ref{t7}) and 
(c)  for the cosmic scale factor (\ref{t9}). Once the time evolution of 
these fields
is known one can determine the time evolution of the energy densities
of the tachyon and matter fields from 
\be
{\rho_\varphi\over\rho_\varphi^0}={V(\varphi)\over \sqrt{1-\dot\varphi^2}},
\qquad 
{\rho_m\over\rho_\varphi^0}=H^2-{V(\varphi)\over \sqrt{1-\dot\varphi^2}}.
\ee{t11}

\section{Results and Discussions}

In order to  solve the two systems of coupled differential
equations for: 
\begin{itemize}
\item[(a)] the non-interacting inflaton field [eqs. (\ref{13}), 
(\ref{14}) and (\ref{15})],
\item[(b)] the non-interacting tachyon field 
[eqs. (\ref{15}), (\ref{t7}) and (\ref{t9})],
\end{itemize}
we have  to choose at first the initial conditions
at $t=0$ (by adjusting clocks). Here we have assumed: 
\begin{itemize}
\item[(i)]  $a(0)=1$ for the dimensionless cosmic 
scale factor,
\item[(ii)]   $H(0)=1$ for the dimensionless Hubble parameter, 
\item[(iii)] the inflaton 
field $\phi(0)$ and the tachyon field $\varphi(0)$  were chosen 
in such a manner that the initial values of the corresponding potential
densities were given by $V(\phi(0))=V(\varphi(0))=1$, 
\item[(iv)]  $\dot\phi(0)=0$ and   $\dot\varphi(0)=0$ so that the initial 
values of the energy densities of the inflaton and tachyon fields refer
only to the corresponding potential densities, and 
\item[(v)] $\varpi(0)=0$ so that  the irreversible processes of
particle production begin just after the time $t=0$.
\end{itemize}

Among several models for the potential density of the scalar field 
(see, for example, Liddle and Lyth~\cite{LLy}) we fix our attention to the 
exponential potential 
\be
\cases{
V(\phi)={\rm exp}[-\mu\phi(t)]+\lambda,\qquad\hbox{for the inflaton field,}\cr
V(\varphi)={\rm exp}[-\mu\varphi(t)]+\lambda,\qquad
\hbox{for the tachyon field},}
\ee{t14}
where $\mu>0$ and $0<\lambda\ll1$ are free parameters, the first one  is 
connected with the slope of the potential
density and has influence on the  
velocity of the field which  rolls down toward the potential minimum, whereas 
the second one represents the asymptotic value of potential
density $V$ and can be interpreted
as  the vacuum energy~\cite{SS,Car,PR,Pad}.

Apart from the two free parameters $\lambda$ and $\mu$
there still remains much freedom to find the solutions of the four 
systems of coupled differential equations, since they do depend on
the parameters: (a)  $w_m$ which is related to the barotropic 
equation of state of the matter field, and
(b) $\alpha$ which  is connected to the transfer of  energy
 from the gravitational field  to the
matter field.

Before discussing the results in detail, let us briefly describe the evolution 
of the system. In the beginning all the energy is in the form of potential 
energy of the scalar field, which has negative pressure and therefore leads to 
an inflationary expansion of the Universe. As the scalar field begins to roll
down its potential $V$, its kinetic energy increases and its potential energy 
decreases, raising its pressure and making its total energy decay. At the same
time, the non-equilibrium pressure $\varpi$ increases in module and the 
irreversible processes of particle production begins. Therefore, the increasing
production of matter and the decay of the potential energy of the scalar field
contribute to slow down the expansion of the Universe and the decelerated 
period begins. During this transition process the non-equilibrium pressure had
started to decrease in module, and a little time after deceleration begins 
$\varpi$ was already driven towards zero. That means that the irreversible
processes are only important during the early Universe, and there is no matter
creation during the decelerated phase and thereafter, so that the matter fields
evolve as usual. Moreover, the roll of the scalar field has been damped by the
friction term ($3H\dot\phi$ or  $3H\dot\varphi$), such that its kinetic energy
tends to zero and its potential energy tends to $\lambda$. As long as 
$\rho_m\gg\lambda$ the Universe evolves as in the standard hot Big-Bang model
and the role of the scalar field is negligible. But when $\rho_m$ became
comparable to $\lambda$ (the vacuum energy of the scalar field), a transition
period takes place and the Universe begins to accelerate.

The time evolution of the acceleration field $\ddot a$ is plotted 
in figure 1 whereas in figure 2 are plotted the time evolution 
of  the  energy densities of inflaton
$\rho_\phi$, tachyon $\rho_\varphi$ and matter $\rho_m$  fields. 
In these figures 
we have chosen the following values for the 
parameters:
$\lambda=0.005$, $\mu=6.5$, $w_m=1/3$, and $\alpha=0.2$. Below
we shall comment how the  changes of these parameters affect the solutions 
of the differential equations. 

\begin{figure}
\begin{center}
\includegraphics[width=7.5cm]{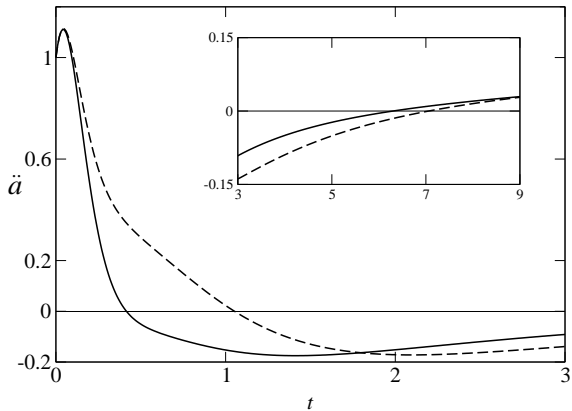}
\caption{Acceleration $\ddot a$  vs time $t$: inflaton field -- 
straight line, tachyon field -- dashed line.}
\end{center}
\end{figure}

We infer from figure 1  that 
there exist three distinct periods for the acceleration field, namely,
(a) an inflationary epoch with an exponential growth to a maximum value 
followed by a decrease  of the acceleration field towards zero, (b) 
a past decelerated period where the acceleration field decreases to 
a maximum negative value followed by an increase towards zero, and 
(c) a present accelerated epoch.  We note 
from figure 1
that the acceleration field of the inflaton begins its decelerated 
period earlier than
the corresponding one for the tachyon.
This behavior is a consequence of the fact that the pressure of the inflaton
field $p_\phi=\dot\phi^2/2-V(\phi)$ may assume positive values, and in fact it
does, due to the increase in the kinetic energy of the inflaton and the 
decrease in its potential energy. Hence, the inflaton field behaves as 
matter with positive pressure for a while, leading to a precocious 
deceleration. By contrast, the pressure of the tachyon field
$p_\varphi=-V(\varphi)\sqrt{1-\dot\varphi^2}$ cannot assume positive values 
(for $V(\varphi)>0$). Its contribution for the deceleration is not as intense
as in the case of the inflaton field, since it can only come close to dust-like
behavior ($p_\varphi\sim 0$). That prolongs the first accelerated period.

We observe from figure 2 that for the energy densities there exist
three distinct epochs which begins with a scalar field dominated period 
followed by a matter field dominated epoch and coming back to a scalar field 
dominated phase.
The following conclusions for the energy densities of the inflaton and 
tachyon fields 
can  be obtained also from  figure 2: 
(a) in the earliest times the energy densities of the
inflaton and tachyon fields coincide, since $\dot \phi^2\ll V(\phi)$ and
$\dot \varphi^2\ll 1$ so that $\rho_\phi\approx V(\phi)\approx V(\varphi)
\approx\rho_\varphi$, (b) during the first accelerated phase the energy density
of the tachyon field decays more slowly than that of the inflaton
field, for two reasons: first, the influence of the downhill roll of the
tachyon field (the increase in $\dot\varphi$) on its respective energy
density is stronger than that of the inflaton field, since the former 
is proportional to $1/\sqrt{1-\dot\varphi^2}$, while the latter is proportional
to $\dot\phi^2$; second, the equation of motion of the tachyon field,
$\ddot\varphi=-(1-\dot\varphi^2)(3H\dot\varphi+V'/V)$, tells us that the change
in $\dot\varphi$ tends to zero as it approaches the value 1, 
therefore $\dot\varphi$ grows more than $\dot\phi$ and stands for a longer 
period nearby the value 1, and
(c) in the transition from the past
decelerated epoch to the present accelerated period the energy density of the 
tachyon becomes smaller than that of the inflaton, 
because the tachyon has rolled faster and further than the inflaton, implying
that $V(\varphi)<V(\phi)$ and therefore $\rho_\varphi<\rho_\phi$ -- since
in this period the potential energies dominate the kinetic energies.

\begin{figure}
\begin{center}
\includegraphics[width=7.5cm]{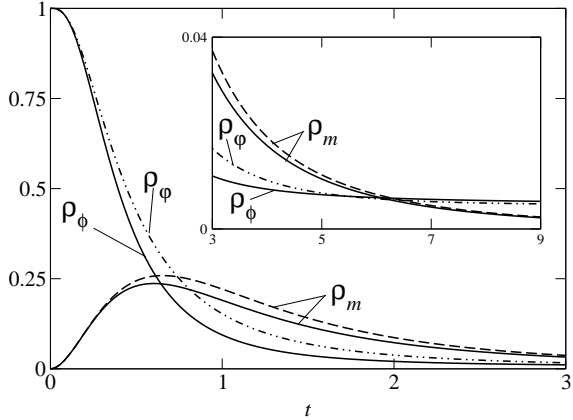}
\caption{Energy densities of inflaton 
$\rho_\phi$, tachyon $\rho_\varphi$
and matter $\rho_m$ fields   vs time $t$: inflaton field -- straight lines, 
tachyon field -- dashed lines.}
\end{center}
\end{figure}

The behavior of the 
matter field can be interpreted as follows: 
the decrease in module of the non-equilibrium pressure $\varpi$ for
the tachyon field is slower -- since its positive acceleration is more 
prolonged -- and therefore more energy is transfered to the matter field than 
the corresponding case for the inflaton field.

We shall comment upon the coefficient that is responsible
for particle creation, namely $\alpha$. By increasing the
value of $\alpha$: (a) the initial accelerated period grows due to
the increase in module of the non-equilibrium pressure, and (b) the 
energy density of the matter field increases so that 
the matter dominated period becomes larger and the present accelerated period 
begins at later times. 

Let us now comment on the coefficient $w_m$ which refers to the barotropic 
equation of state of the matter field. By decreasing its value 
the positive pressure of 
the matter field decreases, and that implies: 
(a) a less pronounced deceleration, 
and (b) a slower decay of the energy density of the matter.

If we decrease the value of the parameter $\lambda$ -- which 
is related to the asymptotic value of the potential density $V$ -- 
we infer that: 
(a) there exists a longer matter dominated period since the vacuum
energy $\lambda$ -- which is the responsible for the present accelerated 
period -- will dominate at later times, and (b) if $\lambda\rightarrow0$ 
there exists no period of present acceleration, since
$\rho_\varphi\approx V(\varphi)\rightarrow0$ and $\rho_\phi\approx 
V(\phi)\rightarrow0$. Hence,
$\lambda$ is  responsible for the present acceleration and can
be interpreted as the dark energy. 

As was previously remarked the parameter $\mu$ is related with the slope 
of the potential density  of the scalar field. By decreasing the value of 
$\mu$ the scalar fields $\phi$ and $\varphi$ roll more slowly so that the decay
of their corresponding energy densities  $\rho_\phi$ and $\rho_\varphi$
is slower and the initial accelerated period becomes larger.
 Moreover, if $\mu$ is very small the decelerated period for the inflaton 
may not exist
due to the fact that the inflaton rolls more slowly when the
slope of its potential density decreases.
This behavior is not followed by the tachyon, since as was previously 
commented $\dot\varphi$
is pulled toward 1 and $\varphi$ rolls more rapidly than $\phi$. Hence, 
even for small values
of $\mu$ there exists a decelerated period for the tachyon field in this model.

As a final remark, we note that  the same general behavior 
of the  acceleration  and of the energy densities fields 
may be obtained by choosing other usual types of
potential densities for the scalar field found in the
literature (see e.g.~\cite{LLy}).

\end{document}